\documentclass[titlepage,12pt]{utarticle}

\usepackage{amsfonts}
\usepackage{amssymb}
\usepackage{epsfig}

\begin{document}

\newcommand{\eL}{{\cal L}}
\newcommand{\half}{\textstyle{1 \over 2}}
\newcommand{\J}{\bf J}
\newcommand{\bP}{\bf P}
\newcommand{\G}{\bf G}
\newcommand{\K}{\bf K}
\newcommand{\M}{{\cal M}}
\newcommand{\bu}{\bf u}
\newcommand{\la}{\lambda}
\preprint{
       DOE-ER-40757-120\\
       HEP-UTEXAS-98-21\\
       }

%------------------------------------------------------------------------

\title{Density Matrices and Geometric Phases for $n$-state Systems}

\author{Luis J. Boya\footnote{Permanent address: Departamento de Fisica Te\'orica, Facultad de Ciencias, Universidad de Zaragoza.  50009 Zaragoza, Spain.}, 
Mark Byrd, Mark Mims and E. C. G. Sudarshan}

\oneaddress{Center for Particle Physics\\
			The University of Texas at Austin\\
			Austin, TX 78712-1081 USA
			\\ {~} \\
			\email{luisjo@posta.unizar.es}\\
			\email{mbyrd@physics.utexas.edu}\\
			\email{mmm@physics.utexas.edu}\\
			\email{sudarshan@physics.utexas.edu}
}

\Abstract{
An explicit parameterization is given for the density matrices for $n$-state systems.  The geometry of the space of pure and mixed states and the entropy of the $n$-state system is discussed.  Geometric phases can arise in only specific subspaces of the space of all density matrices.  The possibility of obtaining nontrivial abelian and nonabelian geometric phases in these subspaces is discussed.
}

\maketitle

%---------------------------------------------------------------------------------------
%\tableofcontents

\pagebreak

%---------------------------------------------------------------------------------------
%---------------------------------------------------------------------------------------
\section{Introduction}

Physical systems with two states or two levels are very well known in physics.  A common 
example is a light beam, which has two polarization states that can exhibit a change in 
phase due to, for example, the geometry of a optical fiber through which it travels \cite{Pan}.  This phase is called a Berry phase or a geometric phase \cite{Berry}; {\it geometric} because it is due to the holonomy in the parameter space\cite{Simon}.  

There are a wide variety of physical applications for three state systems (see for example \cite{DGT}, \cite{3st}, \cite{HioeandEberly}) and more recently $n$-state systems as well (see for example \cite{HioeandEberly}, \cite{Hioe} and \cite{FEL}).  Applications include photon resonances in atomic and molecular systems (see the aforementioned references).  There is interest in the geometry of the spaces of density matrices for quantum statistical mechanics and thermodynamics 
applications as well as ordinary quantum mechanical applications (see \cite{Slater}, \cite{Bloore} and references therein).  

Recently, the particulars of geometric phases for density matrices of three state systems in $SU(3)$ 
representations were studied in Refs. \cite{m1}, \cite{m2}.  There it was discovered that 
nontrivial geometric phases for the pure state density matrices for three state systems occur only in specific cases that can be associated to subgroups within the $SU(3)$ space.  Also given were formulas for the Berry connection one form and curvature two form.  Another, group theoretical, way of deriving these was then pointed out in Ref. \cite{me}.  

One may then ask if there is method of determining, for $n$-state systems, the mixed state density matrices that will exhibit geometric phases and what type of phases we should expect (abelian {\it vs.} nonabelian).  
A general approach to the calculation of geometric phases for mixed state density 
matrices can be found in \cite{Uhlmann}.  

In this paper, the aim is to extend the analyses of the geometric phases present in 2 and 3 state systems to $n$-state systems.
In section \ref{sectionDensityMatrices} we give a brief review of density matrices and parameterize the density matrices, 
{\sl mixed} as well as pure, for an $n$-state system.  Then, in section 
\ref{sectionGeometricPhases} we give a brief history and review of geometric phases.  Then we identify the spaces which could possibly 
give rise to nontrivial geometric phases.  Finally, in section \ref{sectionEntropy}, we briefly discuss the 
entropy\footnote{ 
	The term {\it entropy} used in this paper refers to an ``informational entropy.''  This is not a {\it physical} entropy which would require specification 
	of the particular dynamics of the system.
}
of space of density matrices.   This allows us to measure the ``purity'' of a given density matrix.

%---------------------------------------------------------------------------------------
%---------------------------------------------------------------------------------------
\section{Density Matrices for $n$-state Systems}
\label{sectionDensityMatrices}

Pure states in Quantum Mechanics are represented by one--dimensional subspaces or 
{\sl rays} in Hilbert space, $\cal{H}$.  These states can be characterized as orthogonal 
projections of unit trace
\begin{equation}
\lbrace\hbox{space of pure states}\rbrace\approx
	\lbrace\rho\in\hbox{End}({\cal H})\mid
		\rho^\dagger = \rho,\
		\rho^2 = \rho,\
		\hbox{Tr}\rho = 1
	\rbrace .\label{purestates}
\end{equation}

Von Neumann and Landau originally introduced the notion of {\sl mixed} states into 
Quantum Mechanics \cite{vNeumannA}.  
To include mixed states into our current discussion, 
we recall that it is enough to enlarge (\ref{purestates}) by simply relaxing the 
idempotency condition, $\rho^2 = \rho$ but keeping positivity ($\rho > 0$), thus
\begin{equation}
\lbrace\hbox{space of mixed states}\rbrace\approx
	\lbrace\rho\in\hbox{End}({\cal H})\mid
		\rho^\dagger = \rho,\
		\rho > 0,\
		\hbox{Tr}\rho = 1
	\rbrace .\label{mixedstates}
\end{equation}

As mentioned before, two state systems have been studied in great detail.  
Here we review some of the features that we seek to generalize to systems with $n$ states.  
Consider first the pure states of a 2-state system.  The Hilbert space will be
$\mathbb{C}^2\cong\mathbb{R}^4$.  The unit vectors will form $S^3\subset\mathbb{R}^4$,
and the true {\sl physical} states will be obtained by moding out a phase $S^1$:
\begin{equation}
	S^2=S^3/S^1
		\quad
	\hbox{or}
		\quad
	\beta :S^1\to S^3\to S^2
	\label{hopfBdl}
\end{equation}
which is the fundamental Hopf ($\beta$) fibering.  The $S^1$ is associated to the geometric phase.  

\begin{figure}[h]
\centerline{\epsfxsize=3.0cm \epsfbox{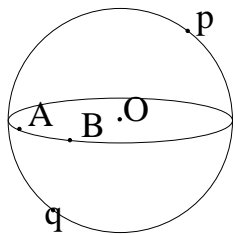}}
\caption{The disk $D^3$ represents the space of mixed states in a 2-state system.
The surface $S^2$ consists of pure states.  The point $p$ and the antipodal point
$q$ form 
an orthogonal pair.}
\label{sphere}
\end{figure}

In the construction
\begin{equation}
	\lbrace\rho\hbox{ pure}\rbrace
		\quad\leftrightarrow\quad
		\hbox{sphere of radius }\half,
\end{equation}
the mixed states can be identified with the {\sl interior} of this 2--sphere
(fig \ref{sphere})
\begin{equation}
	\lbrace\rho\hbox{ mixed}\rbrace
		\quad\leftrightarrow\quad
		\hbox{closed ball }D^3\hbox{ of radius }\half,
\end{equation}
where we have used a radius one for $S^3$ so a radius one-half for $S^2$.  The most mixed state is the center of the ball, $O$, and for any two states $A$ and $B$
in the ball, 
$$
 0 \le \hbox{overlap}(A,B) \equiv |\langle A| B \rangle|^2 \le 1 
$$
is only zero if $B$ is antipodal to $A$.
%figure of sphere...

Following a somewhat similar approach, we will proceed to discuss the geometry of the 
spaces of pure and mixed state density matrices for more general $n$-state systems.

%---------------------------------------------------------------------------------------
\subsection{Pure State Density Matrices}

The space of a pure state density matrix for an $n$-state system is isomorphic to 
$\mathbb{C}$P$^{n-1}$.  This may be seen in two different ways. 

Consider, as above, the common example of a pure state for a $2$-state system. Use the 
space $\mathbb{C}^2 \approx \mathbb{R}^4$, restrict to the transformations that preserve 
the modulus squared, $S^3$, and projects out an overall phase to obtain $\mathbb{C}$P$^1$.  

For a $3$-state system, one would use the space $\mathbb{C}^3 \approx \mathbb{R}^6$,
restricted to the transformations that preserve the modulus squared, $S^5$, and project out an overall phase to obtain $\mathbb{C}$P$^2 \cong S^3/S^1$.  

Now one finds an immediate generalization.  For an $n$-state system, one would use the space 
$\mathbb{C}^n \approx \mathbb{R}^{2n}$, restricted to the transformations that preserve 
the modulus squared, $S^{2n-1}$, and project out an overall phase to obtain 
$\mathbb{C}\mbox{P}^{n-1}$. Therefore {\it the space of pure states for 
an $n$-state system is} 
	\begin{equation}
		\mathbb{C}\mbox{P}^{n-1} =
			{{ \mathbb{C}^n - \{0\} }\over{ \mathbb{C}^1 - \{0\} }} =
			{{ S^{2n-1} }\over{ S^1 }}.
		\label{e.statesA}
	\end{equation}

There is another way to discover this which will lead to a parameterization of the density 
matrices for $n$-state systems.  Consider the density operator for a pure state that is 
represented by an $n \times n$ matrix of zeros with a single $1$ anywhere on the diagonal, {\it viz.},
$$
\left( \begin{array}{crcl}
                   0      &    0    &  \cdots  &    0       \\
                   0      & \ddots  &          &    0         \\
                   \vdots &         &   1      & \vdots       \\
                   0      &    0    & \cdots   &    0         \end{array} \right). 
$$
 
For convenience we may take the $1$ to be in the $n^{\mbox{th}}$ column and $n^{\mbox{th}}$
row. 

Now if we wish to transform this into an arbitrary pure state, $\rho$ transforms as $A \rho A^{-1} = A \rho A^{\dagger}$ under any unitary transformation $A \in U(n)$.  On the other hand, any two pure states are always equivalent under a transformation $A$, {\it i.e.}, $U(n)$ acts transitively on $\mathbb{C}$P$^{n-1}$.  From the previous matrix form, it is clear that the little group is $U(n-1) \times U(1)$ hence
	\begin{equation}
		\mathbb{C}\text{P}^{n-1} = \hbox{Gr}(1,n) =
			{{ U(n) }\over{ U(1)\times U(n-1) }} =
			{{ SU(n) }\over{ U(n-1) }} ,
		\label{e.statesB}
	\end{equation}
where, the Grassmanian of $r$-planes in $\mathbb{C}^n$
	\begin{equation}
		\hbox{Gr}(r,n) =
			{{ U(n) }\over{ U(r)\times U(n-r) }} 
		\label{e.grassman}
	\end{equation}
is well known.

Heuristically, we can go from form (\ref{e.statesA}) to form (\ref{e.statesB}) by
exhibiting the odd-sphere structure\cite{Boya}
\begin{equation}
U(n) \cong S^1\times S^3\ltimes S^5 \ltimes\dots\ltimes S^{2n-1},
\end{equation}
where the product, ($\ltimes$), is a nontrivial (twisted) one.  So
\begin{equation}
		{{ U(n) }\over{ U(1)\times U(n-1) }} =
		\hbox{Gr}(1,n) =
		\mathbb{C}\text{P}^{n-1} =
		{{ S^{2n-1} }\over{ S^1 }}.
\end{equation}

%---------------------------------------------------------------------------------------
\subsection{Mixed State Density Matrices}

Using the pure state operator described above, a parameterization of an $n$-state 
density operator matrix can be given.  Each pure state can be represented by a matrix which 
in diagonal form, would consists of zeros and a single $1$ on the diagonal as above.  Of course 
each class of pure states will have a different nonzero diagonal element.  To achieve a mixture of 
these states, a convex linear combination of them is required.  This linear combination may be 
written as follows:
$$
\rho =  \left( \sum_i a^i \rho_i \right),
$$
where $\rho$ is the mixed state matrix.  The $\rho_i$, $i=1...n$, are the pure state 
matrices and satisfy $\mbox{Tr}(\rho_i \rho_j) = \delta_{ij}$ when normalized.  The $a^i$ satisfy $\sum_i a^i = 1$ as well as the constraint that each is 
positive and between zero and one.  The eigenvalues of this matrix, the $a^i$, can be 
parameterized by the squared components of an $(n-1)$-sphere.  For three state systems the 
eigenvalues are 
$$
a^1 = \cos^2 \phi/2 \; \sin^2 \theta/2 , \;\;\;\;\; a^2 = \sin^2 \phi/2 \; \sin^2 
\theta/2 ,\;\;\;\;\; a^3 = \cos^2 \theta/2.
$$

We can now obtain a general mixed state, one in a generic non-diagonal configuration for 
an $n$-state density matrix.  Let $\rho$ denote the diagonal density operator where the 
diagonal elements are the squared components of the $n-1$ sphere as discussed above and 
let $D$ denote an $SU(n)$ matrix.  Then the mixed 
state density operator matrix is given by
$$
\rho^{\prime} = D \rho D^{-1}.
$$
This parameterization enables one to identify the little group and orbit space of these density matrices.  
One may then investigate which subspaces give rise to nontrivial abelian and non abelian 
geometric phases.  This is the subject of the next section.

%---------------------------------------------------------------------------------------
%---------------------------------------------------------------------------------------
\section{Geometric Phases}
\label{sectionGeometricPhases}

Geometric phases in physics have a long and rich history \cite{snw}, \cite{acw}.  Pancharatnam studied them in connection with photon polarizations over forty years ago \cite{Pan}.  Later Herzberg and Longuet-Higgins showed that the phases can arise in the Born-Oppenheimer approximation describing polyatomic molecules \cite{LH}.  Mead and Truhlar showed that motion of nuclei could be described by an effective Hamiltonian with a ``gauge potential'' to describe the effect of the ``fast'' motion of the electrons on these nuclei to produce the geometric phase \cite{mandt}.  Some time after this Berry rediscovered nontrivial phase factors \cite{Berry} and Simon gave them a geometric description (hence the name, ``geometric phase'') in terms of fiber bundle theory \cite{Simon}.  The classic example is the famous Aharonov-Bohm effect, or the motion of an electron in the field of a magnetic monopole.  Many new interesting topics associated to these phases soon followed.  Among these were analyses in group theoretical terms (\cite{sudgeoph}, \cite{Ali}, \cite{boyageoph}).

Recently geometric phases for $SU(3)$ representations have been investigated in \cite{m1} and \cite{m2} for their connection to three-level systems.  The Lie subgroups of $SU(3)$ were listed there and it was shown that non-zero dynamical phases can only occur in certain particular subspaces of the form $SU(3)/H$ where $H$ is a (Lie) subgroup of $G$.  The space of density matrices of a pure state in a three state system is isomorphic to $SU(3)/U(2) \cong \mathbb{C}$P$^2$.  Coordinates of this manifold were given (see also \cite{me}).  Here, as said, we comment on the possibility of nontrivial phase factors in different subspaces of the space of all density operators for $n$-state systems.

%---------------------------------------------------------------------------------------
\subsection{The Basics}

A deeper treatment of the geometric phase can be found in 
\cite{AA}.  For general  
references see \cite{snw} and \cite{Bohm} in addition to the seminal references given 
above.  Here we sketch the basic concepts needed throughout the section.

Whenever a state in a given quantum mechanical system undergoes a cyclical evolution, any
representative vector will acquire a phase independent of representation.
It is important to realize the resulting overall phase has {\sl two} contributions.
The {\sl dynamical} part depends essentially upon the Hamiltonian (this was well known), 
whereas the {\sl geometrical} part (known in some instances, but rediscovered by
Berry and interpreted by Simon) depends only upon the closed path of evolution, 
hence {\sl geometric}.
This geometric phase is just the holonomy of the loop due to the natural connection in
the $U(1)$-bundle associated to the projection $\{\mbox{vectors}\}\rightarrow\{\mbox{rays}\}$.

In the original derivation by Berry, the motion was adiabatic, and the connection was shown
(by B. Simon) to arise from the effect of the ``slow'' variables on the ``fast'' variables.
Of course, the Lie algebra of the holonomy is generated by the curvature of the connection
(the theorem of Ambrose and Singer, see \cite{KobNom}),
therefore the geometric phase is a gauge invariant concept.

The removal of the unnecessary adiabatic approximation was carried out in \cite{AA}.
They also consider {\sl non-cyclic} evolution with the tacit understanding that the
open path could be closed.  {\it e.g.}, by tracing geodesics from a common point
through the initial and final points.  The path is then closed, and the geodesic segments do
not alter the holonomy.

This naturally leads to a treatment of the effective dynamics of the particular system
as a gauge theory, where   
the symmetries involved depend upon many different aspects
of that particular system; the external parameters used, the adiabatic approximation used, 
etc.  

For a generic path, the holonomy group is $U(1)$.  In the case of degenerate eigenvalues,
the holonomy group can be enhanced to $U(n > 1)$ provided the evolution of the system 
does not remove this degeneracy (as shown in \cite{WandZ}).

The approach we take below is not to try to produce an overall
symmetry group for the theory, but to consider only the little
group of a particular mixed state density matrix.  This will provide a gauge 
group for all matrices in the $SU(n)$ orbit of this state.  We
will then investigate the possible geometric phases within such orbits.  

While this lack of a 
general gauge group for all states in the theory is somewhat 
unsatisfying, this approach turns out to be a very useful one
from the practical point of view.  Physical experiments
often consist of a system prepared in a particular subspace
($SU(n)$ orbit) of the space of all states.  In fact, this 
system often remains within that orbit throughout 
the entire experiment.  It then becomes useful, if you can't
determine the appropriate symmetry group for the entire state
space, to at least determine it for the physically relevant 
subspace.

Notice that this gauge group will, in general, {\sl not} be abelian.  The terms {\it abelian} and {\it
non-abelian geometric phases} then simply refer to the commutativity of the gauge group 
involved.

%---------------------------------------------------------------------------------------
\subsection{Geometric Phases for Mixed States}

We have said that geometric phases depend on the dynamical path 
of a particular system.  The easiest way to determine the possible geometric phases for 
a system is to consider the eigenvalues associated with the effective dynamics
of that system.  For $n$-state systems we note that for different sets of eigenvalues of the system, there exists the possibility of nontrivial abelian and non-abelian geometric phases depending on the particular set of eigenvalues.  This, and the geometry of the spaces of the density matrices, are exhibited in our particular parameterization rather nicely. 

First consider the nontrivial example of a three state system.  When each of 
the three eigenvalues of the mixed state density operator are nonzero and are different, 
the space of transformations of $\rho$ is $SU(3)/(U(1)\times U(1))$, {\it i.e.}, the little group or stability group is $(U(1)\times U(1))$.  
This is a flag manifold 
$ Fl(3) \cong U(3)/U(1)^3 \cong \mathbb{C}\mbox{P}^2 \ltimes \mathbb{C}\mbox{P}^1$.  
One can see that the we may have non trivial geometric phases only of the abelian type.  When the matrix $\rho$ contains 3 nonzero eigenvalues with $2$ identical and one different, the transformation space is $SU(3)/(SU(2)\times U(1))$ and we may possibly observe geometric phases of the nonabelian type as is the case when all eigenvalues are identical.  Of course these spaces exhibit only the possibility of observing geometric phases.  To actually observe them depends on the effective dynamics of the particular system and on the path in parameter space.

For the generic case of $n$-state systems there is an immediate generalization of the argument for 3-state systems.  If we have an $n$-state density operator with all eigenvalues different and nonzero, then the space of transformations is $Fl(n) \cong U(n)/U(1)^n \cong SU(n)/(U(1)\times \cdot \cdot \cdot \times U(1))$, where there are $n-1$ of these $U(1)$ factors.  If two of the eigenvalues are degenerate, there is one factor of $SU(2)$ replacing one of the $U(1)$s.  If three are degenerate, there is a factor of $SU(3)$, etc.  This extends the ideas of \cite{m1} and \cite{m2} from the case of pure 3-state systems to the case of pure and mixed $n$-state systems.  This goes further and offers a scheme for the parameterization of these with $SU(n)$ groups and uses the direct connection between these groups and the transformation spaces of the density operators.

One can now see that the space of mixed state density matrices is locally isomorphic to the following spaces.  When all of the eigenvalues are different the space is $(SU(n)/T^{n-1})\times D^{n-1}$, (see also \cite{Dittmann} for another proof) where the component $D^{n-1}$ is the $(n-1)$ dimensional disc.  This comes from the the parameterization of the diagonal elements in terms of the ($n-1$)-sphere and is topologically a disc but is geometrically an $(n-1)$ dimensional rectangle or rectangular solid.  When there are two degenerate eigenvalues the space is $[SU(n)/(SU(2)\times T^{n-2})]\times D^{n-2}$ and so on.  Here we have used the fact that $T^1\cong U(1)$.

For the case of three state systems we may list 3 possibilities:  all three eigenvalues are different, two are the same and one is different, and all are the same.  Respectively the spaces of their density matrices are locally $(SU(3)/T^{2})\times D^2$, $[SU(3)/(SU(2)\times T^1)]\times D^2$ and a single point.

It is worth noting also that one may arrange the eigenvalues of a diagonalized density matrix such that the identical eigenvalues appear next to each other along the diagonal.  This means that when the matrix is in nondiagonal form, the group transformations can be expressed in a block diagonal form.  These transformations are always allowed since they correspond to a simple change of basis.  In this way we may observe the symmetry breaking associated to the differences in eigenvalue degeneracies.  For instance, if three eigenvalues of a three state system are equal, one can be ``distorted,'' or changed slightly (by an outside influence such as an external magnetic field on a three state molecule).  Then if the population of one of the states is changed, we have broken the symmetry group from $SU(3)$ to $SU(2)\times U(1)$.

%---------------------------------------------------------------------------------------
\section{Entropy}
\label{sectionEntropy}

It is possible to introduce an {\sl entropy} function, $S$, to describe the degree of ``impurity'' of mixed states, $\rho$, in the sense of von Neumann \cite{vNeumannA}.  The definition is 
$$
S(\rho) = - \mbox{Tr}(\rho \ln \rho)
$$
This is a useful definition, even as a thermodynamic function, provided the temperature is high enough in order that the probability of a state will not depend on energy (absence of the Boltzmann factor).  It also gives us a way to identify parts of the space of density matrices as having different ``purities.''  

For $n = 2$ the above definition works rather nicely.  Write
$$
\rho \approx  \left( \begin{array}{crcl}
	\cos^2 \theta/2 &   0             \\
	0               & \sin^2 \theta/2	
        \end{array} \right)  \;\;\;\;\;\;\;\;\;\;  0 \leq \theta \leq \pi/2
$$
where $\approx$ means up to unitary equivalence.  Hence
$$
S[\rho(\theta)] = -(\cos^2 \theta/2)\ln(\cos^2 \theta/2) - (\sin^2 \theta/2)\ln(\sin^2 \theta/2).
$$
As $\theta$ varies from $0$ to $\pi/2$, the entropy varies from $0$ to its maximum $= \ln 2$, and the state varies from a pure one to a generic one to the most mixed one.  Of course, it is enough to compute the entropy for a diagonalized density matrix, since $\mbox{Tr}M = \mbox{Tr} AMA^{-1}$ for any two operators $A,\; M$.

If a diagonal representative of a $3$-state system is 
\begin{equation}
	\rho\cong\left(\begin{matrix} \cos^2 \phi/2 \sin^2 \theta/2 & 0 & 0\\
					0 & \sin^2 \phi/2 \sin^2 \theta/2 & 0\\
					0 & 0 & \cos^2 \theta/2 \end{matrix}
					\right)
	\quad\quad
		0 \le \theta, \phi \le \pi
\end{equation}
the entropy is
$$
S(\rho) = -\mbox{Tr}(\rho \ln \rho) =
			\begin{cases}
				0&  \hbox{for pure states},\\
				0\leq S\leq \ln 3 & \hbox{for generic states}, \\
				\ln 3 &  \hbox{for the most ``mixed'' state}.
			\end{cases}
$$

The generalization to an $n$-state system is obvious, with similar limits
$$
S(\rho) = -\mbox{Tr}(\rho \ln \rho) =
			\begin{cases}
				0&  \hbox{for pure states},\\
				0\leq S\leq \ln n &\hbox{for generic states}, \\
				\ln n &  \hbox{for the most ``mixed'' state}.
			\end{cases}
$$
Finally, notice that, for $n > 2$, states with the same entropy are {\sl not} equivalent under $SU(n)$, because the rank is higher than one and the orbits on the adjoint irreducible representation need {\sl more} than one label.

%---------------------------------------------------------------------------------------
%---------------------------------------------------------------------------------------
\section{Conclusions and Comments} 

We have shown that the mixed state density matrices for $n$-state systems can be parameterized in terms of squared components of an ($n-1$)-sphere and unitary matrices.  This lets one immediately identify the little groups and therefore orbits of the space of density matrices for particular sets of eigenvalues.  These little groups can be seen as structure that allows the possibility of geometric phases for the system.  Thus when a system described by an $n$-state density matrix undergoes a change in the physical parameter space (the orbit space), the system may exhibit a geometric phase with a ``gauge group'' that corresponds to the little group of the space.  Given the parameterization presented here, the identification of the little group is transparent.

Using this parameterization of the density matrices gives one a way of obtaining an {\sl explicit} parameterization of the pure and mixed state density matrices that is amenable to calculations.  One may parameterize the unitary matrices for $SU(n)$ in terms of Euler parameters as in \cite{me}.  There the Euler parameters for $SU(3)$ were given in detail.   Using this parameterization of $SU(3)$ in ref. \cite{me2} we can calculate explicit geometric phases for the case of $3$-state systems.  This is the subject of a forthcoming paper.

This parameterization also lets us describe the geometry of the spaces of density matrices simply in terms of well known geometrical objects.  The entropy function is now parameterized in terms of these squared components of the $(n-1)$-sphere which describes the diagonal elements (eigenvalues) of a mixed state density matrix.  The purity of states can then be described using this function and the normals to the constant entropy surfaces describe the directions of maximum entropy increase.  This is an area we are currently investigating.

%---------------------------------------------------------------------------------------
%---------------------------------------------------------------------------------------
\section{Acknowledgments} 

One of us (M. S. B.) would like to thank Prof. Duane Dicus whose help and support enabled the completion of this paper and Paul B. Slater for pointing out some of the references above and for useful comments.  We also thank Eric Chisolm for helpful comments.  This research was supported in part by the U.S. Department of Energy under Contract No. DE-EG013-93ER40757.
Also, one of us (L. J. B.) thanks the Spanish CAICYT for partial support.

\end{document}